\def\equationautorefname~#1\null{Equation (#1)\null}
\def\sectionautorefname~#1\null{Section #1\null}
\def\subsectionautorefname~#1\null{Section #1\null}
\def\subsubsectionautorefname~#1\null{Section #1\null}
\def\figureautorefname~#1\null{Figure #1\null}

\documentclass[twocolumn]{aastex63}
\usepackage{lineno}
\usepackage{amsmath}
\usepackage{xparse}
\usepackage{hyperref}
\usepackage{longtable}
\usepackage[flushleft]{threeparttable}
\usepackage[titletoc,title]{appendix}
\usepackage{bm}

\newcommand*{\dc}{--''--}
\newcommand*{\ar}{\autoref}
\newcommand*{\ct}{\citet}
\newcommand*{\ctp}{\citep}
\newcommand*{\cta}{\citealp}

\received{}
\revised{}
\accepted{\today}
\graphicspath{{./}{figures/}}
\begin{document}

\title{Opacity of the highly ionized lanthanides and the effect on the early kilonova} 

\correspondingauthor{Smaranika Banerjee}
\email{smaranikab@astr.tohoku.ac.jp}

\author[0000-0002-0786-7307]{Smaranika Banerjee}
\affiliation{Astronomical Institute, Tohoku University, Aoba, Sendai 980-8578, Japan}

\author[0000-0001-8253-6850]{Masaomi Tanaka}
\affiliation{Astronomical Institute, Tohoku University, Aoba, Sendai 980-8578, Japan}
\affiliation{Division for the Establishment of Frontier Sciences, Organization for Advanced Studies, Tohoku University, Sendai 980-8577, Japan}

\author{Daiji Kato}
\affiliation{National Institute for Fusion Science, National Institutes of Natural Sciences, Oroshi-cho, Toki, Gifu 509-5292, Japan}
\affiliation{Department of Advanced Energy Engineering Science, Kyushu University, Kasuga, Fukuoka 816-8580, Japan}

\author{Gediminas Gaigalas}
\affiliation{Institute of Theoretical Physics and Astronomy, Vilnius University, Saulėtekio av. 3, LT-10257 Vilnius, Lithuania}

\author[0000-0003-4443-6984]{Kyohei Kawaguchi}
\affiliation{Institute for Cosmic Ray Research, The University of Tokyo, 5-1-5 Kashiwanoha, Kashiwa, Chiba 277-8582, Japan;
Center for Gravitational Physics, Yukawa Institute for Theoretical Physics, Kyoto University, Kyoto, 606-8502, Japan}

\author[0000-0002-7415-7954]{Nanae Domoto}
\affiliation{Astronomical Institute, Tohoku University, Aoba, Sendai 980-8578, Japan}

\begin{abstract}
We investigate the effect of the presence of lanthanides ($Z = 57 - 71$) on the kilonova
at $t\sim$ hours after the neutron star merger for the first time.
For this purpose, we calculate the atomic structures and the opacities for selected lanthanides:
Nd ($Z = 60$), Sm ($Z = 62$), and Eu ($Z = 63$).
We consider the ionization degree up to tenth (XI), applicable for the ejecta at
$t \sim$ a few hours after the merger, when the temperature is $T \sim 10^{5}$ K.
We find that the opacities for the highly ionized lanthanides are exceptionally high,
reaching $\kappa_{\rm exp}\sim 1000\,\rm cm^{2}\,g^{-1}$ for Eu, due to the highly dense energy levels.
Using the new opacity, we perform radiative transfer simulations to show that the early light curves become fainter by a (maximum)
factor of four, in comparison to lanthanide-free ejecta at $t\sim 0.1$ day.
However, the period at which the light curves are affected is relatively brief
due to the rapid time evolution of the opacity in the outermost layer of the ejecta.
We predict that for a source at a distance of $\sim 100$ Mpc, UV brightness for lanthanide-rich ejecta shows a drop to
$\sim 21 - 22$ mag at $t\,\sim 0.1$ day and the UV peaks around $t\sim0.2$ day with a magnitude of $\sim19$ mag.
Future detection of such a kilonova by the existing UV satellite like \textit{Swift} or the upcoming UV satellite ULTRASAT
will provide useful constraints on the abundance in the outer ejecta and the corresponding nucleosynthesis conditions
in the neutron star mergers.

\end{abstract}

\keywords{stars: neutron, nucleosynthesis, r-process, opacity, radiative transfer, gravitational waves}

\section{Introduction} \label{sec:intro}

\begin{figure*}[t]
  \begin{tabular}{c}
 
   \begin{minipage}{0.5\hsize}
      \begin{center}
        \includegraphics[width=\linewidth]{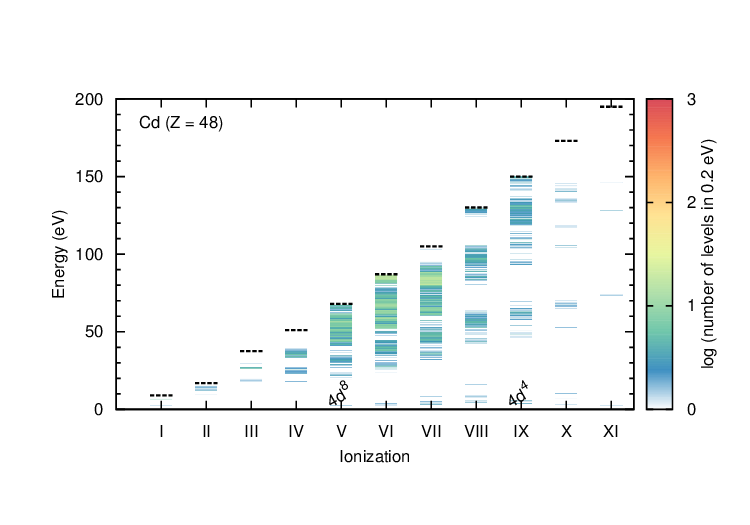}
      \end{center}
    \end{minipage}
    
    \begin{minipage}{0.5\hsize}
      \begin{center}
        \includegraphics[width=\linewidth]{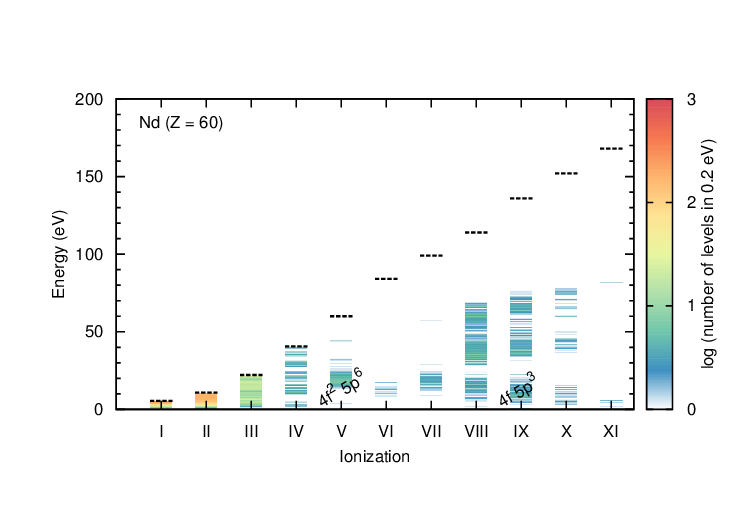}
      \end{center}
    \end{minipage}
    \\
    \begin{minipage}{0.5\hsize}
      \begin{center}
        \includegraphics[width=\linewidth]{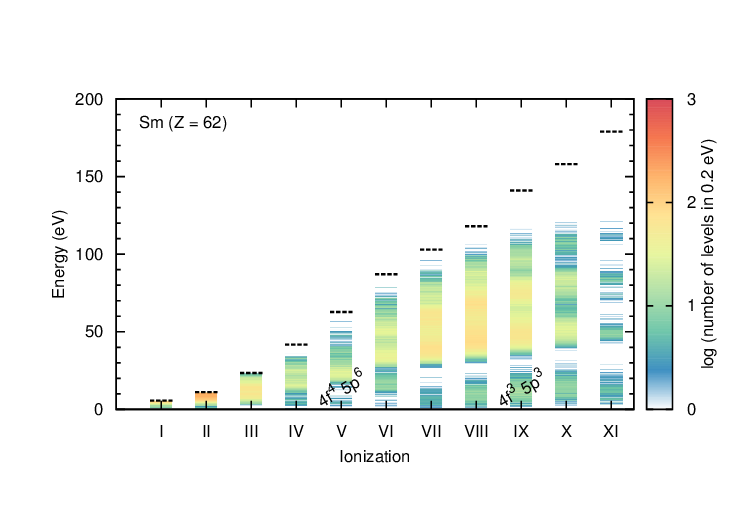}
      \end{center}
    \end{minipage}
    
     \begin{minipage}{0.5\hsize}
      \begin{center}
        \includegraphics[width=\linewidth]{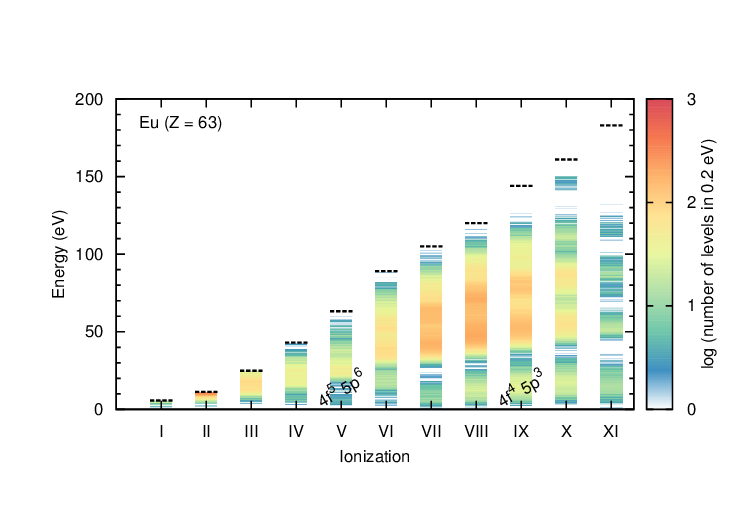}
      \end{center}
    \end{minipage}
  
\end{tabular}
  \caption{The energy level distribution for three lanthanide elements Nd, Sm, and Eu
    compared with that of a light $r$-process element Cd. The dashed black lines show the potential energies for each ionization.}
  \label{fig:elev}
\end{figure*}

\begin{figure*}[t]
  \begin{tabular}{c}
 
   \begin{minipage}{0.5\hsize}
      \begin{center}
        \includegraphics[width=\linewidth]{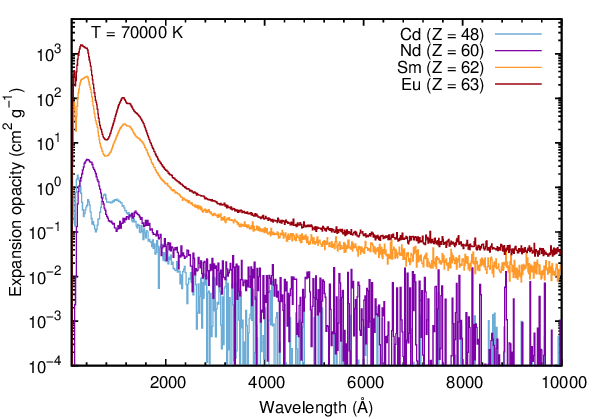}
      \end{center}
    \end{minipage}
    
     \begin{minipage}{0.5\hsize}
      \begin{center}
        \includegraphics[width=\linewidth]{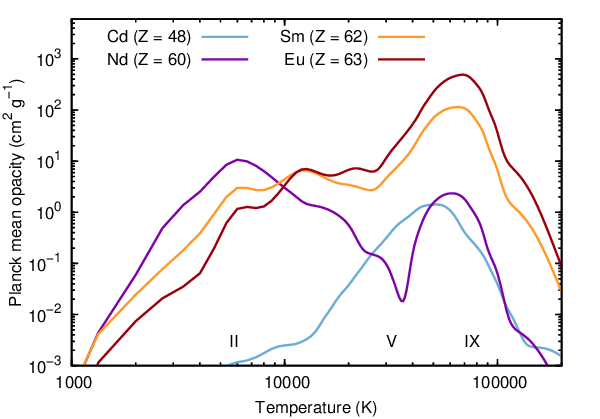}
      \end{center}
      \end{minipage}
      \end{tabular}
  \caption{The expansion opacity as a function of wavelength at $T\,=\,70000$ K (left panel) and
    the Planck mean opacity as a function of temperature (right panel) at $\rho \,=\, 10^{-10}\, \rm g\,cm^{-3}$,
    and $t = 0.1$ day for lanthanides Nd, Sm, and Eu. Opacity of the light $r$-process element Cd is also shown for comparison \ctp{Smaranikab20}.
  }
  \label{fig:expop}
\end{figure*}
It has long been hypothesized that the heavy elements are synthesized in the neutron star mergers
(e.g., \cta{Lattimer74, Eichler89, Freiburghaus99, Korobkin12, Wanajo14}).
The radioactive decays of the freshly synthesized heavy elements give rise to a transient in the
ultraviolet, optical, and near-infrared wavelengths, called kilonova (e.g., \citealt{Li98, Kulkarni05, Metzger10}).
In fact, such a kilonova AT2017gfo (e.g., \cta{Coulter17, Yang17, Valenti17}) from the neutron star merger 
has already been observed by the follow up observations of the gravitational wave event GW170817 \citep{Abbott17a},
confirming the neutron star mergers to be a site of the $r$-process nucleosynthesis.

The light curve of AT2017gfo was bright in UV and optical band at the epoch of detection
(e.g., \cta{Coulter17, Yang17,Valenti17}).
The light curve evolved to be fainter at optical and brighter at NIR in the timescale of $t \sim$ a week
(e.g., \cta{Cowperthwaite17, Smartt17, Drout17, Utsumi17}).
The late time ($t > 1$ day) light curve is well explained by kilonova
(e.g., \ctp{Kasen17, Tanaka17, Shibata17, Perego17, Rosswog18, Kawaguchi18}). 
However, the origin of the early emission ($t<1$ day) has not reached a consensus \citep{arcavi18}.

Several models exist for the early kilonova.
For example, the early emission might be powered by the radioactive decays of the heavy elements,
similar to the later phase \ctp{Waxman18, Villar17, Smaranikab20}.
Alternatively, the early emission might result from the interaction of the relativistic jet with the surrounding ejecta
\citep{Kasliwal17, Piro18, Nativi21, Klion21} 
or by $\beta$-decay of the free-neutrons \ctp{Metzger15}.
It is important to understand the early kilonova because the early emission can reveal the abundance in the outer ejecta
because photons only from the outer layer can escape at an early time.
Hence, understanding the kilonova starting from an early time is crucial to understanding the
abundance pattern from the outer to the inner ejecta.

The major uncertainty in modelling the early kilonova comes from the lack of the detailed opacity of the $r$-process elements.
The shape of the kilonova light curve is mainly determined by the opacity in the ejecta (e.g., \cta{Metzger10, Kasen13, Tanaka13}).
Hence, modelling the kilonova requires detailed opacity.
Past studies have shown that the bound-bound transitions contribute the most
to the opacity in the kilonova \ctp{Kasen13, Tanaka13, Fontes15, Fontes20, Wollaeger17, Tanaka18, Tanaka20a}.
Calculations of the bound-bound opacity require the atomic data, which are largely unavailable for the early kilonova.
This is due to the fact that, at an early time, the elements are highly ionized, maximum up to tenth ionization
(or XI in spectroscopic notation; hereafter, the spectroscopic notation is used to describe the ionization)
at $T\,\sim\,10^5$ K, the typical temperature at $t\sim 0.1$ day. 
Most of the earlier studies have performed the atomic calculations only up to the ionization IV,
which are suitable for the opacity at $t\geq 1$ day.
Hence, to derive opacity at the early time, the atomic structure calculations for the highly ionized heavy elements are necessary.

\ct{Smaranikab20} performed the atomic calculations for the light $r$-process elements at the ejecta condition
suitable at the early time.
However, their study does not include lanthanides (elements with $Z = 57 - 71$).
Lanthanides are the elements with open $4f$-shell. 
Previous studies on the opacity at $t\geq 1$ day have shown that the presence of lanthanides can significantly affect
the opacity and the light curve.

In this paper, we perform the first atomic opacity calculation for the selected lanthanides:
Nd ($Z = 60$), Sm ($Z = 62$), and Eu ($Z = 63$) up to the ionization XI and study the impact on the early kilonova emission.
We show our new atomic and opacity calculations in \ar{sec:op}.
Using the new opacity, we perform radiative transfer simulations for lanthanide-rich ejecta in neutron star mergers
in \ar{sec:radtr}.
The validity of standard method of opacity calculation in the expanding media (expansion opacity formalism, \cta{Sobolev60})
for the highly ionized lanthanides is discussed in \ar{subsec:sob}.
We also investigate the future prospects to detect lanthanide-rich kilonova in \ar{subsec:fu_pr}.
Finally, we summarize our conclusions in \ar{sec:conclusion}. AB magnitude system is adopted throughout the article.

\section{Opacity in neutron star merger}\label{sec:op}
In the neutron star merger ejecta, different processes such as electron scattering, free-free, bound-free,
and the bound-bound transitions contribute to the opacity.
Here we calculate the different opacity components for the selected lanthanides,
Nd ($Z = 60$), Sm ($Z = 62$), and Eu ($Z = 63$) following \ct{Smaranikab20}.
The ejecta conditions are assumed to be suitable for the early time, i.e., the density of the ejecta is taken to be $\rho = 10^{-10}\,\rm g\,cm^{-3}$,
which is the typical density at $t \sim 0.1$ day for an ejecta mass of $M_{\rm ej}\sim 0.01M_{\odot}$
and the elements are considered to be ionized up to $\sim$ XI corresponding to the typical temperature of $T \sim 10^{5}$ K. 

First, we estimate the electron scattering, free-free, and the bound-free opacities for lanthanides.
We find that the electron scattering and the free-free opacity for lanthanides (ionized up to $\sim$ XI) are
$\sim 3\,\times10^{-2} \,\rm{cm^2\,g^{-1}}$ and $\sim 2\,\times10^{-3}\,\rm{cm^2\,g^{-1}}$, respectively,
which are not very different from those for light $r$-process elements.
Note that we assume the electron temperature is the same as the ejecta temperature.
Also, we confirm that the bound-free opacity is not important in our chosen wavelength range ($\lambda = 100\,-\,35000 \rm \AA$).
This is because the fraction of the photons with energy beyond the ionization potential is not significant.
A similar conclusion was made for the early bound-free opacity for the light $r$-process elements \ctp{Smaranikab20}.

Next, we explore the bound-bound opacity for lanthanides, for which the atomic structure calculations are necessary.
Since the complete atomic data calibrated with experiments are unavailable for the highly ionized lanthanides,
we perform the theoretical atomic structure calculations as described in the following section.

\subsection{Atomic structure calculation} \label{sec:atomic}

\subsubsection{Method}\label{subsec:at_meth}
We use HULLAC (Hebrew University Lawrence Livermore Atomic Code, \cta{Bar-shalom01}) for the atomic calculations.
HULLAC uses fully relativistic orbitals to calculate the energy levels and radiative transition probabilities.
A set of orbital functions is obtained by solving the single electron Dirac equation with a parametric, central field potential,
which includes both nuclear field and the spherically averaged electron-electron interaction.
The central field potential for a given electron charge distribution can be obtained by solving the
Poisson equation with the boundary condition that the potential converges to $(Z-q)/r$ at $r = \infty$
for an element with atomic number $Z$ and the number of the electrons $q$.
The nuclear charge is assumed to be a point one ($Z\delta(r)$).
The charge density distribution of the electrons is expressed as \citep{Bar-shalom01}:
\begin{equation}\label{eqn:e_distr}
  \rho(r) = -4\pi r^{2} q N [r^{l+1}\rm {exp}^{-\alpha r/ 2}]^{2},
\end{equation}
where $N$ is the normalization factor and $\alpha$ is a free parameter related to the average radii ($<r>$)
of the Slater type orbital as $(2l + 3)/<r>$. 
The free parameter $\alpha$ is obtained by minimization of the first order configuration averaged energies
for selected configurations.

The all-electron zero-order solution or the configuration state function (CSF) is the
anti-symmetrized products of the orbitals in a chosen coupling scheme ($jj$ coupling in this case).
After constructing the zero-order wave function,
the magnetic and the retardation effect of the interaction from the other electrons (Breit term)
and the quantum electrodynamic energy correction are taken into account.
Finally, the atomic wave function is constructed using the linear combination of the CSFs.

We perform the atomic calculations including the excited configurations for the opacity.
Such calculations require the ground configurations.
This is because HULLAC solves the Dirac equation with a certain central potential
derived by the electron distribution in the ground state.
However, the ground configurations are not well established for the
highly ionized lanthanides (ionization $\geq$ V).
Therefore, for the three lanthanides (Nd, Sm, and Eu) ionized to the degree V - XI,
we use the ground configurations estimated with HULLAC.
The detailed strategy of the calculations to estimate the ground configurations is provided in Appendix A.
For the all the other ions, we use the ground configurations as provided in NIST atomic spectra database (\cta{NIST20}).
Using the ground configurations together with the excited configurations, the atomic calculation for the opacity is performed.
We provide all the configurations used in \ar{tab:confg}.
The ground configurations and all the configurations used for minimization are shown in bold.

It is noted that assuming the central potential has a parametric form (\ar{eqn:e_distr}) in HULLAC
can be a potential source of uncertainty in atomic calculations.
Different atomic codes adopt different approaches:
for example, Flexible Atomic Code (FAC, \cta{Gu08}) determines the potential in a self-consistent manner.
Nevertheless, comparison of atomic data calculated by HULLAC and more ab-initio GRASP2018 code \ctp{Grasp19}
shows good match for various ions with ionization $<$ V (e.g., \cta{Tanaka18, Gaigalas19, Rynkun21}).
Similar comparison for highly ionized lanthanides remains within the scope for the future work.

\begin{figure*}[t]
  \begin{tabular}{c}

  \begin{minipage}{0.5\hsize}
      \begin{center}
       
       \includegraphics[width=\linewidth]{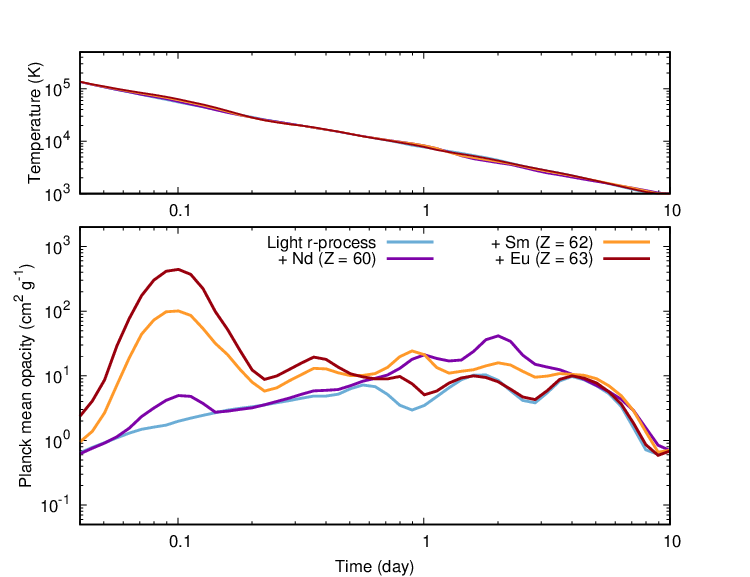}
        
      \end{center}
    \end{minipage}
  
   \begin{minipage}{0.5\hsize}
      \begin{center}
        
        \includegraphics[width=\linewidth]{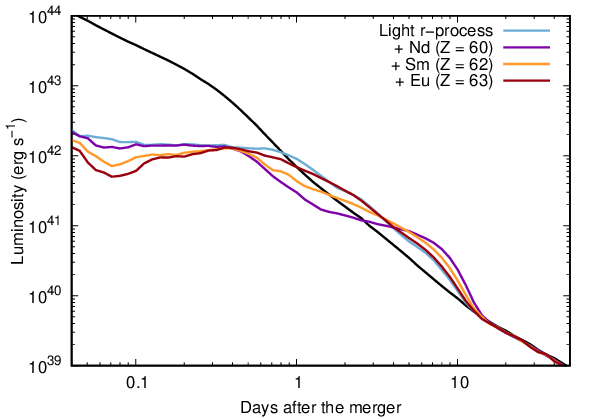} 
        
      \end{center}
    \end{minipage}

      \end{tabular}
  \caption{\textbf{Left panel:} The temperature (upper panel) and the Planck mean opacity (lower panel)
    evolution with time at outer layer of the ejecta ($v$ $\sim$ 0.19c).
    The ejecta are assumed to be composed of one lanthanide element (Nd or Sm or Eu with $X_{\rm La} = 0.1$) and the
    light $r$-process elements. The opacity evolution for the ejecta containing only the light $r$-process elements
    is also shown for comparison (blue curve, \cta{Smaranikab20}).
    The presence of lanthanides causes a dramatic increase in the opacity in the early time at $t \sim 0.1$ day.
    \textbf{Right panel:} The deposited luminosity (black curve) and the bolometric luminosity (colored curves).
    The early bolometric luminosity shows a drop if lanthanides are present.}
  \label{fig:lcla}
\end{figure*}
\subsubsection{Energy level distribution}\label{subsubsec:elev_sec}

The energy level distribution obtained from our atomic calculations is shown as a function of ionization in \ar{fig:elev}.
The color scale represents the level density in the 0.2 eV energy bin.
Since our main purpose is to calculate the bound-bound opacity, we show the energy levels only below the ionization threshold.
Also, the energy level distribution of the light $r$-process element Cd ($Z = 48$, \cta{Smaranikab20}) is shown for comparison.

The energy level distribution of lanthanides is denser in comparison with the light $r$-process element Cd (\ar{fig:elev}).
This behavior is common over all the ionizations (for the low-ionized cases,
see \cta{Kasen13, Tanaka13, Fontes15, Fontes20, Wollaeger17, Tanaka18, Tanaka20a}).
For highly ionized lanthanides, this trend is more prominent in the cases of the Sm and Eu than Nd. 
This is due to the large number of available states for the electrons (high complexity measures)
in the middle of lanthanide series.

For a given element, the density of the energy levels increases with the ionization.
The energy distribution is densest around the ionization of VII - IX for lanthanides.
For instance, for Eu, the number of energy levels increases significantly from the ionization VI - VIII.
This is caused by the presence of the two open shells
($4f$- and $5p$-shell, see \ar{tab:confg}) in highly ionized lanthanides.
We can understand this by comparing the ground configurations of Eu VI and VIII.
Eu VI has only one open shell in the ground configurations ($4f^{4}\,5p^{6}$, \ar{tab:confg}),
whereas Eu VIII has two open shells ($4f^{4}\,5p^{4}$, \ar{tab:confg}).
The same is true for the excited configurations in both ions.
Hence, Eu VIII has a higher energy level density than Eu VI.

For a given ionization, the energy level distribution varies depending on the elements.
This is due to the difference in the number of $4f$-electrons (\ar{tab:confg}).
For example, Eu IX has four $4f$-electrons in the ground configuration ($4f^{4}\,5p^{3}$, \ar{tab:confg}),
whereas Nd IX has only one ($4f\,5p^{3}$, \ar{tab:confg}).
As a result, Eu IX shows a higher energy level density than Nd IX (\ar{fig:elev}).


\subsection{bound-bound opacity} \label{sec:bb_op}
Using the new atomic data, we calculate the bound-bound opacity for the three highly ionized lanthanides (Nd, Sm, and Eu).
In an expanding media such as that in the cases of supernova and kilonova,
the opacity is calculated by adopting the expansion opacity formalism \citep{Karp77}.
To calculate the expansion opacity, the contribution from the multiple lines is averaged within a chosen wavelength bin.
The strength of the individual transition is calculated by assuming Sobolev approximation \citep{Sobolev60}. 

In the expanding media, the photons are continuously redshifted,
causing the photons to progressively coming into resonance with different lines.
Note that the resonance occurs over a certain wavelength range rather than at a particular wavelength because of the intrinsic profiles of the atomic lines. In neutron star merger ejecta, the line profile is predominantly determined by the thermal motion. If the thermal widths of the lines are negligible in comparison with the line spacing,
we can evaluate the strength of each transition by the Sobolev optical depth, which is not dependent on the line profile function. More on the validity of the Sobolev approximation is discussed in \ar{subsec:sob}.

When the Sobolev approximation is valid, the expansion opacity is calculated as follows.
If there are $N$ strong lines inside an arbitrarily chosen wavelength bin of $\Delta \lambda$,
the velocity gradient required to redshift the photons from one line to another is given as:
\begin{equation}\label{eqn:vel_grad}  
  \Delta v = c\dfrac {\Delta \lambda/ N} {\lambda}.
\end{equation} 
Such a velocity gradient corresponds to a mean free path of $\Delta v t$ at a time $t$.
The corresponding absorption coefficient within the wavelength bin of $\lambda$ to $\lambda + \Delta \lambda$
is written as \ctp{Kasen13}:
\begin{equation}\label{eqn:aexp}
\alpha_{\rm exp}(\lambda) = \dfrac{1}{\Delta v t} =\dfrac{1}{ct}\dfrac{\lambda}{ \Delta \lambda} N.
\end{equation}
In this expression, only strong lines are considered.
To include the contribution from the weak lines, a modified version derived by \ct{Eastman93} is used:
\begin{equation}\label{eqn:alphaexp}
    \alpha_{\rm{exp}}(\lambda) = \frac{1}{ct}\sum_{l}\frac{\lambda_{l}}{\Delta \lambda}(1 -e^{-\tau_{l}}),
\end{equation}
where $\lambda_{l}$ is the transition wavelength in a chosen wavelength interval of $\Delta \lambda$.
The Sobolev optical depth at the transition wavelength ($\tau_{l}$) is calculated as
\begin{equation}\label{eqn:tau}
    \tau_{l} = \frac{\pi e^{2}}{m_{\rm{e}} c} n_{l}\lambda_{l}f_{l}t,
\end{equation}
where $n_{l}$ is the number density of the lower level of the transition, and $f_{l}$ is the oscillator strength of the transition. 
Then, we can calculate the expansion opacity as the absorption coefficient per unit mass density:
\begin{equation}\label{eqn:kexp}
    \kappa_{\rm{exp}}(\lambda) = \dfrac{\alpha_{\rm exp}(\lambda)}{\rho}.
\end{equation}
Using this formalism, we calculate the opacity for a single element ejecta with the density $\rho = 10^{-10}\,\rm g\,cm^{-3}$
at $t \sim$ 0.1 day.

We assume local thermodynamic equilibrium (LTE) to calculate the ionization fraction of the elements by solving the Saha ionization equation and to determine the population of the excited levels via Boltzmann statistics.

The low density of the neutron star merger ejecta ($\rho \sim 10^{-10}\, \rm g\, cm^{-3}$ even at $t\sim0.1$ day)
is not enough to establish LTE via collisional processes alone.
Nevertheless, LTE can be established via radiative transitions in the optically thick regions inside the photosphere,
especially in the early time when most of the ejecta are optically thick.

LTE might not be valid if the non-thermal processes from the radioactive decay significantly affect ionization and excitation.
However, \ct{Kasen13} find that the ratio of the non-thermal to thermal excitation rate at $t \sim 1$ day,
when the radioactive power released per particle is $\sim 1\, \rm eV\,s^{-1}$ and the typical transition energy is $\sim 1$ eV
at the temperature $T \sim 5000$ K is negligible (the ratio is $\sim 10^{-8}$).
Extending the calculation to early time at $t\,\sim 0.1$ day, when the radioactive power released per particle is
$\sim 100\, \rm eV\,s^{-1}$ \ctp{Metzger10} and the typical transition energy can be as high as $\sim 10$ eV at the temperature $T \sim 10^5$ K,
we find the ratio is not significant (a rough estimate shows the ratio is about $\le 10^{-8}$).
A similar argument can be made for the non-thermal ionization at the early time.
Hence, it is expected that the non-thermal processes do not make the system largely deviated from LTE at the timescale of interest.
At a later time, when the ejecta become less dense and more transparent, larger deviation from LTE is expected
(for more discussion on non-LTE opacity, see \cta{Pognan22}).

The expansion opacity for lanthanides are exceptionally high (left panel of \ar{fig:expop}).
For instance, the expansion opacity at its peak reaches $\kappa_{\rm exp}\sim 1000\,\rm cm^{2}\,g^{-1}$ for Eu at $T\sim70000$ K.
On the other hand, the opacity of the light $r$-process element Cd can reach only up to $\kappa_{\rm exp}\sim 1\,\rm cm^{2}\,g^{-1}$
under the same condition. This is due to the significantly higher number of the energy levels in the highly ionized lanthanides (\ar{fig:elev}).

The expansion opacities show a strong wavelength dependence, with a higher value at shorter wavelengths (left panel of \ar{fig:expop}).
This is caused by the larger number of transitions at the shorter wavelengths. 
Moreover, the opacity for lanthanides show distinct peaks at short wavelengths
(e.g., see $\lambda\sim 500\, \rm \AA$ and $\lambda\sim 1200 \, \rm \AA$ in \ar{fig:expop}),
which is due to the fact that there are numerous strong transitions at these wavelengths.

The temperature dependence of the expansion opacity is estimated by convolving it with the blackbody function to calculate
the Planck mean opacity (see right panel of \ar{fig:expop}).
The Planck mean opacities for different elements show distinct peaks at temperatures $T \sim 5000$ K and $T \sim 70000$ K.
At the high temperature, Eu has the maximum opacity among the other two lanthanides.
On the other hand, the opacity for Nd is highest at low temperature.

The opacity at high temperature reflects the density of energy levels. 
At the temperature $T \sim 70000$ K, where the opacity peaks appear, lanthanides are ionized to $\sim$ VII - IX.
At this ionization range, the level density of lanthanides is the highest (\ar{sec:atomic}).
We argue that, at high temperature, relatively high energy levels contribute to the opacity.
This is in contrast with the lower ionized (i.e., low temperature) case,
where mostly the lower lying energy levels are important (see \cta{Tanaka20a}).
This is because, at low temperature, only the low lying levels are populated (by Boltzmann distribution).
However, at high temperature, even the relatively higher lying energy levels can be populated,
making the transitions between the high energy levels possible.
Hence, the density of the levels in a wider energy range is important at high temperature.

It is worth mentioning that the opacity is affected by the completeness of the atomic data
as our results show that even higher lying levels are important for the opacity at high temperatures.
Hence, we investigate whether our atomic data include essential transitions
(i.e., whether our atomic data are sufficiently complete for the opacity).
We find that the atomic data are mostly complete for opacity. More details can be found in Appendix B.

\begin{figure*}[t]
  \begin{tabular}{c}
 
   \begin{minipage}{0.5\hsize}
      \begin{center}
        \includegraphics[width=\linewidth]{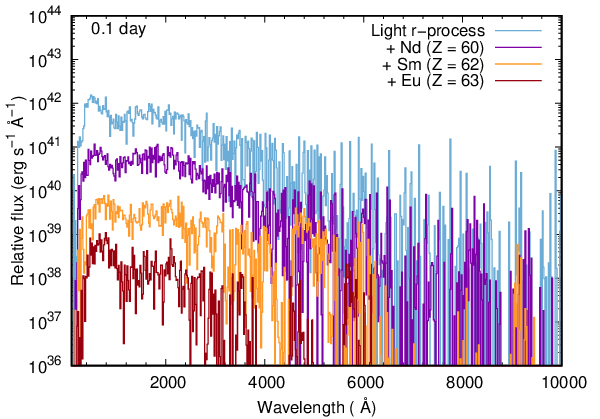}
      \end{center}
    \end{minipage}
    
     \begin{minipage}{0.5\hsize}
      \begin{center}
        
        \includegraphics[width=\linewidth]{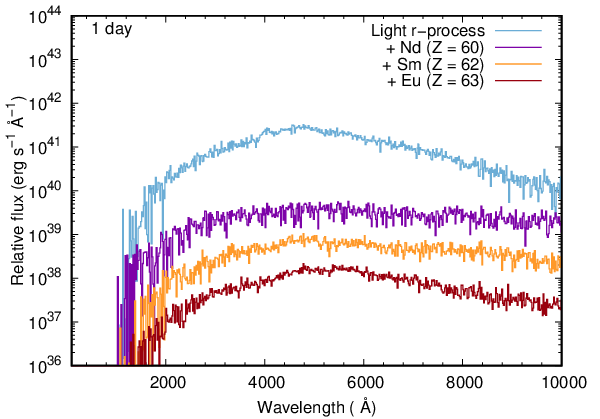}
      \end{center}
      \end{minipage}
      \end{tabular}
  \caption{Scaled spectra for different models at $t = 0.1$ day (left panel), $t = 1$ day (right panel).
  The scaling factors are 1000, 100, 10, and 1 for lanthanide-free ejecta, ejecta containing Nd, Sm, and Eu, respectively.}
  \label{fig:spec}
\end{figure*}


\section{Radiative transfer simulation} \label{sec:radtr}
\subsection{Model}\label{subsec:model}
In neutron star merger, the heaviest elements are produced mainly in the tidal ejecta component,
i.e., in the ejecta that is mostly distributed towards the equatorial plane (e.g., \cta{Bauswein13, Just15, Sekiguchi15, Kullmann22, Just22}).
Hence, the kilonova observed in the equatorial direction is likely to show the effect of the presence of lanthanides.
We calculate the light curve for such a kilonova from a neutron star merger using a time- and wavelength-dependent Monte Carlo radiative transfer code \citep{Tanaka13,Tanaka17, Kawaguchi18}.
The code calculates the multi-color light curves and spectra for a given a density structure and electron fraction
($Y_{\rm{e}}$) distribution assuming the homologously expanding motion of the ejecta.
The radioactive heating rate of $r$-process nuclei is calculated according to
$Y_{\rm{e}}$, by using the results from \citet{Wanajo14}.
The code adopts a time-dependent thermalization factor from \citet{Barnes16}.
Our simulation considers the wavelength range $\lambda \sim 100 - 35000\, \rm \AA$. 

We adapt a spherical ejecta model \citep{Metzger10} with a power-law density structure $\rho \propto r^{-3}$
with a velocity range of $v\,=$ 0.05c to 0.2c and a total ejecta mass of $M_{\rm ej}\, =\,0.05M_{\odot}$
(same as the fiducial model of \cta{Smaranikab20}).
The abundance in the ejecta is assumed to consist of the single lanthanide element
(Nd, Sm, or Eu, considered as different models) with a mass fraction $X_{\rm La} = 0.1$.
Such a lanthanide fraction is obtained in an ejecta with $Y_{\rm{e}}\,=$ 0.20,
the typical value for $Y_{\rm{e}}$ for the equatorial ejecta. The remainder of the ejecta are considered to have the light $r$-process abundance.
The abundance for the light $r$-process elements is determined by using the results from \citet{Wanajo14}
for $Y_{\rm{e}}\,=$ 0.30 $-$ 0.40. A flat mass distribution is considered for each value in the $Y_{\rm{e}}$ range.
We renormalise the abundance to match the total mass fraction of the light $r$-process elements to be $0.9$.
Note that we assume the heating rate is only from the light $r$-process elements because including single lanthanide does not change the heating rate significantly.

Here we mention that performing the radiative transfer simulation using the complete linelist is not feasible since the number of transitions is extremely high.
For instance, the linelist for Eu can consist up to $\sim 0.3$ billion lines (\ar{tab:confg}).
In contrast, the total number of lines for the light $r$-process abundance is $\sim 10$ million.
Hence, we make a reduced linelist for lanthanide elements Sm and Eu for the ionization $>$ V.
For this purpose, we randomly select the transitions from the original linelist by keeping the statistical properties the same.
The detailed scheme and the validity are discussed in Appendix C.

\subsection{Results}\label{sec:lbol}

\ar{fig:lcla} (right panel) shows the bolometric luminosities for different models.
We find that at $t \sim 0.1$ day, the bolometric luminosities for lanthanide-rich ejecta
($L_{\rm bol}\sim 0.5-1 \times10^{42}\,\rm erg\,s^{-1}$, different for different models) is fainter than lanthanide-free ejecta
($L_{\rm bol} \sim 2\times10^{42}\,\rm erg\,s^{-1}$) by a factor of four to two, depending on the models.
The light curves for lanthanide-rich ejecta rise afterwards and show no difference with that of lanthanide-free case at $t \sim 0.4$ day.
Finally, the luminosities for lanthanide-rich models drop and show deviation again at around $t \sim 1$ day.

The shape of the early bolometric light curve is determined by the opacity in the outermost layer in the ejecta ($v \ge 0.19c$).
This is because, in the early time, the diffusion sphere lies at the outermost layer of the ejecta.
At $t = 0.1$ day, the temperature of the outermost layer provides the suitable condition ($T < 70000$ K)
to reach the ionization range of VII - IX, where the opacity peaks appear for lanthanides (left panel of \ar{fig:lcla}).
Such a rise in the opacity in the outermost layer causes the luminosity to drop at $t \sim 0.1$ day.

As the ejecta expands further, the temperature in the outer layer decreases,
crossing the temperature (and ionization) range where the opacities peak.
This results in the luminosity rising at $t \sim 0.4$ day. Finally, at around $t \sim 1$ day,
the outermost layer of the ejecta cools down enough ($T < 10000$ K)
so that the opacity peaks of the low-ionized lanthanides appear (see left panel of \ar{fig:lcla}). 
Hence, the luminosities drop again at $t \sim 1$ day.
Hence, we see that the extremely high opacities for the highly ionized lanthanides affect the light curves 
only for a brief period of time ($t \leq 0.4$ day), reflecting the rapid temperature evolution.

Different lanthanides show the different extent of drops in the luminosity at different times.
For instance, at $t= 0.1$ day, the luminosity of Eu-rich ejecta is the most affected, 
whereas at $t = 1$ day, the drop is the most significant for the ejecta containing Nd.
The extents of the drop are determined by the peak opacities for lanthanides at different temperatures (right panel of \ar{fig:expop}).
For instance, at high temperature, i.e., the condition at $t= 0.1$ day, the opacity is maximum for Eu.
In contrast, at a low temperature, i.e., the condition at $t= 1$ day, Nd has the maximum opacity.
This explains the light curve shapes in the presence of the different lanthanides.

\ar{fig:spec} shows the typical spectra in the presence of lanthanides at $t= 0.1$ and $t= 1$ days.
The spectra are almost featureless.
Since the accuracy of our atomic data is not enough and we use the reduced linelist, which affects the detailed spectral feature,
we do not attempt to discuss the individual elemental signature.
Instead, we note that the spectra rapidly evolves from UV to the optical wavelength range from $t= 0.1$ to $t= 1$ days,
consistent with the expectation. The same trend is observed in all models. 

\begin{figure*}[t]
  \begin{tabular}{c}  
    
    \begin{minipage}{0.5\hsize}
      \begin{center}
        \includegraphics[width=\linewidth]{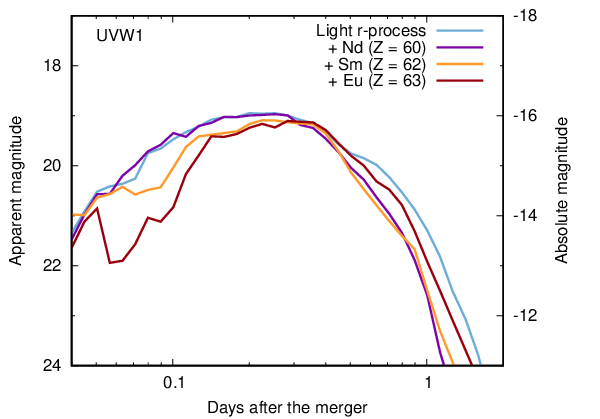}
      \end{center}
    \end{minipage}
   
     \begin{minipage}{0.5\hsize}
      \begin{center}
        
        \includegraphics[width=\linewidth]{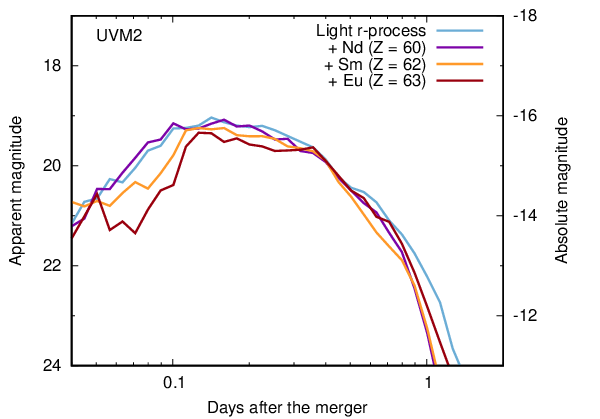}
      \end{center}
    \end{minipage}
  
 \\
\begin{minipage}{0.5\hsize}
      \begin{center}
        \includegraphics[width=\linewidth]{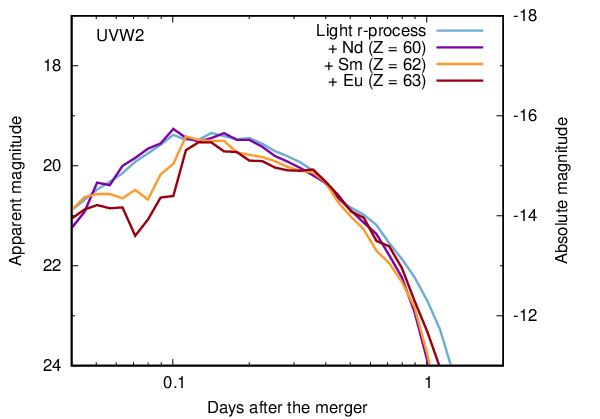}
      \end{center}
    \end{minipage}

\end{tabular}
  \caption{Comparison of UV magnitudes between different models for a source at a distance 100 Mpc.
    The magnitudes are shown for three \textit{Swift} filters UVW2, UVM2, and UVW1
    with the mean wavelengths 2140 $\rm \AA$, 2273 $\rm \AA$, and 2688 $\rm \AA$, respectively \ctp{Roming05}.
  The magnitudes drop at $t \sim 0.1$ day if the lanthanides are present in the ejecta.
  }
  \label{fig:mag}
\end{figure*}


\section{Discussions}\label{sec:discussion}
Our work shows that the opacities for the highly ionized lanthanides are exceptionally high owing to the extremely dense energy levels.
Moreover, we show the luminosity is suppressed in the early time in the presence of lanthanides in the ejecta.
In this section, we discuss the validity of the expansion opacity formalism for the highly ionized lanthanides.
We also discuss the future detection prospects for the early kilonova from lanthanide-rich ejecta.

\subsection{Validity of the Sobolev approximation}\label{subsec:sob}
The expansion opacity at a chosen wavelength interval of $\Delta \lambda$ is derived by taking the cumulative contribution
from all the lines inside $\Delta \lambda$ with the assumption that there is sufficient space between the strong lines.
If the intrinsic line profiles overlap, i.e., if the line spacing ($\Delta \lambda_{\rm line} = \Delta \lambda / N$)
is comparable to the thermal width of the line ($\Delta \lambda_{\rm th}$), such treatment can not represent the true opacity.
Since our calculations show the high number of transitions caused by the high density of the lines (\ar{sec:atomic}),
we check whether the line profiles do not overlap and if the Sobolev approximation is valid.

By following \ct{Kasen13}, we define a critical opacity when the thermal width of the line is equal to the line spacing,
i.e., $\Delta \lambda_{\rm th} = \Delta \lambda/ N = \Delta \lambda_{\rm line}$.
Under such condition, the velocity required to redshift the photon between two consecutive lines (\ar{eqn:vel_grad}) can be given by the thermal velocity $v_{\rm th}$ :
\begin{equation}\label{eqn:vth}
v_{\rm th} = c\dfrac{\Delta \lambda_{\rm th}}{\lambda} = c\dfrac{\Delta \lambda}{\lambda}\dfrac{1}{N}.
\end{equation}
The value of the $v_{\rm th}$ is $\sim 4\rm{km\,s^{-1}}$ at the typical temperature $T\sim 10^{5}$ K at the time $t\sim 0.1$ day
using an average atomic mass number of $A = 150$.
Using \ar{eqn:vth}, \ar{eqn:alphaexp}, and \ar{eqn:kexp}, we can derive the critical opacity as follows: 
\begin{equation}\label{eqn:kcrit}
\begin{split}
    & \kappa_{\rm{crit}}   = \frac{1}{\rho v_{\rm th}t} \\ & \sim 3\, \rm {cm^{2}\,g^{-1}} \bigg(\dfrac{\rho}{10^{-10}\, \rm {g\,cm^{-3}}}\bigg)^{-1} \bigg(\dfrac{\textit{v}_{\rm th}}{4\,\rm{km\,s^{-1}}}\bigg)^{-1} \bigg(\dfrac{\it t}{\rm 0.1\, \rm {d}}\bigg)^{-1}.
\end{split}
\end{equation}
The critical opacity is independent of the wavelength, while it depends on the temperature and density of the ejecta at a particular time.
If the expansion opacity exceeds the critical opacity, the intrinsic line spacing is smaller than the line width,
i.e., the lines overlap with each other.
Then the expansion opacity using the Sobolev approximation for the radiative transfer can not represent the true
opacity in the expanding media.

At a time $t \sim$ 0.1 day after the neutron star merger, for a density $\rho \sim 10^{-10}\,\rm g\,cm^{-3}$,
and a temperature of $T\, \sim \, $ 70000 K, $\kappa_{\rm{crit}} = 3\,\rm cm^{2}\,g^{-1}$ as shown in \ar{eqn:kcrit}.
The expansion opacity, under the same condition, can reach up to $\kappa_{\rm exp} \sim 1000\,\rm cm^{2}\,g^{-1}$
at far-UV ($\lambda\,\le\,2000\,\rm \AA$, left panel of \ar{fig:expop}), exceeding the value of the critical opacity.
Hence, using the expansion opacity at $t\sim0.1$ day for lanthanides can not represent the true opacity at far-UV.
Consequently, the light curves are possibly affected in the far-UV wavelengths.
Nevertheless, our calculation is most likely to remain unaffected at $\lambda \ge 2000 \rm \, \AA$,
which is the detection range of the existing UV instruments like \textit{Swift} \ctp{Roming05}.
The alternative treatment of opacity calculation and its implication to the early kilonova light curve will be discussed in future work.

\subsection{Future prospects}\label{subsec:fu_pr}
  
In this section, we discuss the prospects of observing an early kilonova from lanthanide-rich ejecta.
\ar{fig:mag} shows the magnitudes in the three different \textit{Swift} UVOT filters \ctp{Roming05}
for a source at 100 Mpc for different models. Our results show that the UV brightness for lanthanide-rich ejecta drops at $t\,\sim 0.1$ day,
reaching $\sim 21-22$ mag depending on models. The brightness increases afterwards, reaching $\sim 19$ mag at $t\sim0.2$ day.
Finally, the brightness decreases to $>22$ mag after $t\sim1$ day.

The extents of drop and the slope of the light curves are different for the different models.
For instance, at $t\,\sim 0.1$ day, the magnitudes for Eu-rich ejecta are the faintest,
whereas Nd-rich ejecta show faintest magnitudes at $t\,\sim 1$ day.
Moreover, the presence of Eu makes the light curve rise faster at $t\,\sim 0.1$ day,
whereas the presence of the Nd makes the light curve fall faster at $t\,\sim 1$ day.
This is because of the differences in the opacity in the outermost layer in the presence of the different lanthanides,
as discussed in \ar{sec:lbol}.

The early UV signals for lanthanide-rich ejecta are bright enough to be detected by \textit{Swift}
(with a limiting magnitude of 22 mag for an exposure time of 1000 s, \cta{Roming05}),
provided the kilonova is discovered early enough so that the prompt observation can be started.
Such a kilonova is also a good target for the upcoming wide-field UV satellite ULTRASAT
(limiting magnitude of 22.4 mag for 900 s of integration time, \cta{Sagiv14}).
Future detection of such a kilonova will provide the clues to the abundance pattern in the outer layer of the ejecta,
which can give useful constraints on the nucleosynthesis condition in the neutron star merger. 
Moreover, detecting a rapid rise at $t\,\sim 0.1$ day in UV is likely to be an indicator of the presence of lanthanide element with a similar property to Eu.
On the other hand, detection of a rapid decline at $t\,\sim 1$ day in UV will likely to point that Nd-like lanthanide element is present in the ejecta.

We note that our calculation is limited to provide the trustable result only for the observation at the
$\lambda \ge 2000\, \rm \AA$ at $t\sim0.1$ day.
This is because we calculate the opacity using the expansion opacity formalism which not valid
at $\lambda \le 2000\, \rm \AA$ at $t\sim0.1$ day (\ar{subsec:sob}).

\section{Conclusions}\label{sec:conclusion}
To investigate the early kilonova emission from lanthanide-rich neutron star merger ejecta,
we perform the atomic opacity calculation for the three lanthanides Nd ($Z\,= 60$), Sm ($Z\,= 62$), and Eu ($Z\,= 63$).
For the atomic calculation, we consider the ionization up to XI,
which is the maximum ionization at a typical condition of $T\,\sim \,10^{5}$ K at $t= 0.1$ day.
Our opacity calculations with the new atomic data show that lanthanide opacity can be exceptionally high,
reaching $\kappa_{\rm exp}\,\sim 1000 \,\rm{cm^{2}\,g^{-1}}$ for Eu (left panel of \ar{fig:expop}),
due to the dense energy levels in the highly ionized lanthanides (\ar{fig:elev}).

Using the new opacity, we perform the radiative transfer simulations to calculate the early kilonova
from lanthanide-rich neutron star merger ejecta.
Our models assume that the abundance of the ejecta is the mixture of the single lanthanide
(Nd, Sm, or Eu with a fraction of $X_{\rm La} = 0.1$) and the light $r$-process elements.
Such a lanthanide-rich kilonova may replicate the ejecta condition for a kilonova observed at equatorial direction.

We find that in the presence of lanthanides, the bolometric light curves show a brief period of
luminosity drop at $t \sim 0.1$ day by a (maximum) factor of four in comparison to lanthanide-free case.
The luminosities rise to the same value as lanthanide-free case at $t\sim0.4$ day and finally drops again at $t \sim 1$ days (right panel of \ar{fig:lcla}). 
The shape of the light curve is determined by the opacity in the outermost layer in the ejecta.
The opacity there changes as the the temperature (ionization) changes with the expansion of the ejecta (left panel of \ar{fig:lcla}).
The extents of the luminosity drop are different depending on lanthanide element present since the maximum opacity are different for different lanthanides (\ar{sec:bb_op}).

The UV light curves show the same trends as the bolometric light curves. 
For a source at 100 Mpc the UV brightness drops to $\sim 21-22$ mag at $t\,\sim 0.1$ day (\ar{fig:mag}).
The brightness increases afterwards to reach $\sim 19$ mag at $t\sim0.2$ day, beyond which,
the brightness decreases to $>22$ mag around $t\sim1$ day (\ar{fig:mag}). 
The extents of drop and the slopes of the light curves are different for the different models.
We show that it is possible to detect the early kilonova even for lanthanide-rich ejecta by \textit{Swift}
\ctp{Roming05}, if the kilonova is discovered early enough.
Also, such a kilonova can be detected using the upcoming wide-field UV satellite ULTRASAT \ctp{Sagiv14}.
Detection of such a kilonova in the early time will provide the abundance pattern in the outer ejecta and
can put constraints on the nucleosynthesis condition in the neutron star mergers.

We note that our spherical ejecta model with a homogeneous abundance pattern is designed to represent
the kilonova observed in the equatorial direction is simple.
In reality, the ejecta structure is likely to be more complicated.
For example, many simulations predict the presence of a faster moving outer layer (e.g., \cta{Kyutoku14}).
In such cases, although the unique features in the light curves will still be present,
the significance and the timescale might be different from those observed in our models.
Exploring such possibilities in the future is of interest.
Moreover, it is emphasized that our models can provide the trustable light curves only at $\lambda \geq 2000\, \rm \AA$.
This is because we use the expansion opacity which is not justified in the far-UV wavelengths ($\lambda \leq 2000\, \rm \AA$)
if the highly ionized lanthanides are present.
The alternative opacity treatment and the effect on the early kilonova will be explored in future work.

\acknowledgments

Numerical simulations presented in this paper were carried out with
Cray XC-B at the Center for Computational Astrophysics,
National Astronomical Observatory of Japan;
and at the computer facility in the Yukawa Institute for Theoretical Physics (YITP),
Kyoto University, Japan. The authors would like to thank the anonymous reviewer for his/her comments,
which helped to improve the manuscript. SB wants to thank S. Saito for useful discussion.
This research was supported by the Grant-in-Aid for Scientific Research
from JSPS (19H00694,20H00158,21H04997) and MEXT (17H06363). 
\vspace{5mm}


\bibliography{bib_folder.bib}{}
\bibliographystyle{aasjournal}
\renewcommand{\thetable}{\Alph{section}\arabic{table}}
\renewcommand\thefigure{\thesection\arabic{figure}} 

\appendix

\section{Ground configurations of lanthanides}\label{appendix:A}
In this section, we describe our strategy to estimate the ground configurations
for the highly ionized lanthanides.
In the highly ionized lanthanides ($\geq$ V), the outermost shells are either $4f$- or $5p$-shell.
Hence the electrons are removed either from $4f$- or $5p$-shell for further ionization.
NIST atomic spectra database (NIST ASD, \cta{NIST20}) provides the ground configurations for highly ionized lanthanides
assuming $5p$-electron is removed for further ionization in most of the cases for $\geq V$.
However, such results are based on the relatively simplified theoretical calculations with different approximations
\ctp{Carlson70, Rodrigues04, Sugar75a, Martin78}. Thus, we calculate the ground configurations of highly ionized lanthanides using HULLAC.

We prepare different test cases for the atomic calculations in the following way.
HULLAC calculates the atomic orbitals by solving the Dirac equation with a central potential,
which is determined and optimized based on the given electron distribution. 
We perform the atomic calculations for various central potentials: (1) calculated by changing the electron distribution in $4f$ and $5p$ orbitals 
and (2) optimizing for energy levels belonging to different set of configurations.
For each case, we identify the CSF with the largest mixing coefficient for the lowest energy level.
Then, the configuration generating the CSF of the ground level is taken to be the ground configuration. 
If a certain configuration appears to be the ground one in all the calculations for a particular ion,
we regard it as the ground configuration. 

We explain our strategy by taking Eu as an example. The ground configuration of Eu IV as suggested by NIST ASD is $\rm 4f^{6}\, 5p^{6}$
(experimentally verified).
We test whether the ground configuration of Eu V are $4f^{5}\, 5p^{6}$ (as provided in the NIST ASD) or $4f^{6}\, 5p^{5}$,
corresponding to the $4f$ and $5p$ electron removal from Eu IV ion.
As the total number of electrons in $4f$ and $5p$ orbitals is 11, the effective potential
on a single electron is constructed based on the distribution of 10 electrons.
If the ground configuration is given as $4f^{5}\, 5p^{6}$, the single electron potentials are constructed based on the
electron distributed either as $4f^{5}\, 5p^{5}$ or $4f^{4}\, 5p^{6}$. Then we optimize the potential for $4f^{5}\, 5p^{6}$.
We denote these two cases as Case A and Case B in \ar{tab:config_test_euv}.
Similarly, if the ground configuration is $4f^{6}\, 5p^{5}$, the potential can be constructed for the electrons distributed
either as $4f^{6}\, 5p^{4}$ or $4f^{5}\, 5p^{5}$, and the potential can be optimized for $4f^{6}\, 5p^{5}$. These are denoted as Case C and Case D.
The energy levels are calculated for both candidate configurations in all the cases.
The atomic calculations with these four different cases show that the ground state always belongs to the configuration $4f^{5}\, 5p^{6}$.  
Therefore, we regard $4f^{5}\, 5p^{6}$ as the ground configuration for Eu V.

Similar calculations for the other ions of Eu show the convergence of the ground configurations for most of the ions
(shown in bold in \autoref{tab:config_test_eu}). 
For some ions, however, the results of the four cases do not converge. An example is Eu VII as shown in \autoref{tab:config_test_euvii}.
In such a case, we perform another set of calculation by employing both candidate configurations for the energy minimization
(Cases $\rm A^{'}$, $\rm B^{'}$, $\rm C^{'}$ in \autoref{tab:config_test_euvii}).
If the ground configurations converge in these cases ($\rm A^{'}$, $\rm B^{'}$, $\rm C^{'}$), we choose that configuration as the ground configuration.
In the case of Eu VII, $4f^{4}\, 5p^{5}$ is the ground configuration in all the Cases $\rm A^{'}$, $\rm B^{'}$, $\rm C^{'}$,
and hence, it is regarded as the ground configuration.
The ions which require these additional calculations are indicated with asterisk in \autoref{tab:config_test_eu}.
The same strategy is applied to evaluate ground configurations of other highly ionized lanthanides (Nd ($Z$ = 60) and Sm ($Z = 62$)).
All the ions reach convergence either after Cases A-D or Cases $\rm A^{'}$, $\rm B^{'}$, $\rm C^{'}$.

The ground configurations obtained from HULLAC are different from those provided in the NIST ASD,
which shows that, $5p$-electron removal starts from ion V for further ionization.
In contrast, our results show that $5p$-electron removal starts from ionization VI or VII, depending on the elements.
For ionization VI and higher, NIST ASD adopts the results from the theoretical calculations by \ct{Carlson70},
which provides the ground configuration by removing the consecutive least bound electron.
The least bound orbitals are determined from the solution of the relativistic Hartree-Fock wavefunction for the neutral atoms.
On the contrary, we calculate atomic energy levels for individual ions with an effective central
field potential by taking electron-electron interaction into account.
Hence, we choose to use the ground configurations obtained from the HULLAC instead of the NIST ASD
for the final calculation for the opacity including excited confiurations (\ar{tab:confg}).


\begin{table}[t]
   \begin{threeparttable}
\centering
\caption{Summary of ground configuration calculation for Eu for ionization V - XI}
\label{tab:config_test_eu}
\begin{tabular}{@{}lllllll@{}}
\hline
\textbf{Ionization} & $\mathbf{5p^{6}}$ & $\mathbf{5p^{5}}$ & $\mathbf{5p^{4}}$ & $\mathbf{5p^{3}}$ & $\mathbf{5p^{2}}$ & $\mathbf{5p}$ \\ 
\hline
V          & $\mathbf{4f^{5}\,5p^{6}}$        &  $4f^{6}\,5p^{5}$        &          &          &          &          \\
VI         & $\mathbf{4f^{4}\,5p^{6}}$        &   $4f^{5}\,5p^{5}$       &          &          &          &          \\
VII        & $4f^{3}\,5p^{6}$        &   $\mathbf{4f^{4}\,5p^{5*}}$       &          &          &          &          \\
VIII       & & $4f^{3}\,5p^{5}$                 &   $\mathbf{4f^{4}\,5p^{4}}$   &                &          &          \\
IX         &    &    &$4f^{3}\,5p^{4}$      &   $\mathbf{4f^{4}\,5p^{3}}$           &          &          \\
X          &    &    &   & $4f^{3}\,5p^{3}$       &   $\mathbf{4f^{4}\,5p^{2}}$                            &          \\
XI         &    &    &    &   & $4f^{3}\,5p^{2}$& $\mathbf{4f^{4}\,5p}$                              \\ 
\hline
\end{tabular}
\begin{tablenotes}
    \item[] The result for ground configurations are shown in bold for each ionization
    \item[] The configurations with asterix requires two configurations for optimization
    
\end{tablenotes}
 \end{threeparttable}
\end{table}


\begin{table*}[t]
   \begin{threeparttable}
\centering
\caption{The list of configurations used in different strategies for Eu V}
\label{tab:config_test_euv}
\begin{tabular}{@{}lllll@{}}
\hline
\textbf{Cases } & \textbf{Potential on the} & \textbf{Optimization } & \textbf{Ground }  \\
& \textbf{single electron}&&\textbf{configuration}\\
\hline

A          & $4f^{5}\,5p^{5}$      &  $4f^{5}\,5p^{6}$              &      $4f^{5}\,5p^{6}$           \\

B         & $4f^{4}\,5p^{6}$        &    $\dc$        &            $4f^{5}\,5p^{6}$  \\ 

C         & $4f^{6}\, 5p^{4}$       &  $4f^{6}\,5p^{5}$               &      $4f^{6}\, 5p^{6}$           \\

D         & $4f^{5}\, 5p^{5}$ & $\dc$   & $4f^{6}\, 5p^{6}$\\
\hline
\end{tabular}
\begin{tablenotes}
    \item[] The energy levels are calculated for the configurations $4f^{5}\,5p^{6}$ and $4f^{6}\,5p^{5}$
    
\end{tablenotes}
 \end{threeparttable}
\end{table*}


\begin{table*}[t]
   \begin{threeparttable}
\centering
\caption{The list of configurations used in different strategies for Eu VII}
\label{tab:config_test_euvii}
\begin{tabular}{@{}lllll@{}}
\hline
\textbf{Cases } & \textbf{Potential on the} & \textbf{Optimization } & \textbf{Ground }  \\
& \textbf{single electron}&&\textbf{configuration}\\
\hline
A & $4f^{3}\,5p^{5}$        & $4f^{3}\,5p^{6}$         & $4f^{3}\,5p^{6}$  \\
B & $4f^{2}\,5p^{6}$        & $\dc$                   & $4f^{3}\,5p^{6}$  \\   
C & $4f^{4}\, 5p^{4}$        & $4f^{4}\,5p^{5}$         & $4f^{4}\, 5p^{5}$  \\
D & $4f^{3}\,5p^{5}$          & $\dc$                   & $4f^{4    }\,5p^{5}$ \\   
$\rm A^{'}$           & $4f^{3}\,5p^{5}$   &  $4f^{3}\,5p^{6}$ & $4f^{4}\,5p^{5}$           \\
&& $4f^{4}\,5p^{5}$   & \\
$\rm B^{'}$        & $4f^{2}\,5p^{6}$      &  $\dc$           &      $4f^{4}\,5p^{5}$           \\ 
$\rm C^{'}$       & $4f^{4}\,5p^{4}$       &  $\dc$           &      $4f^{4}\,5p^{5}$           \\
\hline
\end{tabular}
\begin{tablenotes}
    \item[] The energy levels are calculated for the configurations $4f^{3}\,5p^{6}$ and  $4f^{4}\,5p^{5}$
    
\end{tablenotes}
 \end{threeparttable}
\end{table*}

\clearpage
\begin{center}
\begin{longtable}{lllll}
\caption{The summary of the HULLAC calculations. The ground configurations and the configurations used for optimization are shown in bold.
The last column shows the number of energy levels below the ionization threshold.}\\

\hline

\textbf{Ion} & \textbf{Configurations} & \textbf{$N_{\rm level}$ } & \textbf{$N_{\rm line}$ } &\textbf{$N_{\rm line}^{*}$ }\\
\hline 
\endfirsthead
\multicolumn{3}{l}{\tablename\ \thetable\ -- \textit{Continued from previous page}} \\
\hline

\textbf{Ion} & \textbf{Configurations} & \textbf{$N_{\rm level}$ } & \textbf{$N_{\rm line}$ } &\textbf{$N_{\rm line}^{*}$ }\\
\hline
\endhead
\hline  
\multicolumn{3}{r}{\textit{Continued on next page}} \\
\endfoot
\hline
\endlastfoot
Nd I & \begin{tabular}{l} ${\bf 4f^{4} 6s^{2}}$, ${\bf 4f^{4} 6s 5d}$, ${\bf 4f^{4} 6s 6p}$, ${\bf 4f^{4} 6s 7s}$, ${\bf 4f^{4} 6s 8s}$,\\ $4f^{3} 5d 6s^{2}$, $4f^{3} 5d^{2} 6s$, $4f^{3} 5d 6s 6p$\end{tabular} & 12215 & 11784660 & 37121\\ 
Nd II & \begin{tabular}{l} ${\bf 4f^{4} 6s}$, ${\bf 4f^{4} 5d}$, $4f^{3} 5d^{2}$, $4f^{3} 5d 6s$, $4f^{4} 6p$,\\ $4f^{3} 5d 6p$, $4f^{3} 6s 6p$\end{tabular} & 6888 & 3947992 & 2281283\\ 
Nd III & \begin{tabular}{l} ${\bf 4f^{4}}$, $4f^{3} 5d$, $4f^{3} 6s$, $4f^{3} 6p$, $4f^{2} 5d^{2}$,\\ $4f^{2} 5d 6s$, $4f^{2} 5d 6p$, $4f^{2} 6s 6p$\end{tabular} & 2252 & 458161 & 225413\\ 
Nd IV & \begin{tabular}{l} ${\bf 4f^{3}}$, $4f^{2} 5d$, $4f^{2} 6s$, $4f^{2} 6p$, $4f 5d^{2}$,\\ $4f 5d 6s$, $4f 5d 6p$\end{tabular} & 474 & 23864 & 15982\\ 
Nd V & \begin{tabular}{l} ${\bf 4f^{2} 5p^{6}}$, $4f^{3} 5p^{5}$, $4f 5p^{6} 7p$, $4f 5p^{6} 6s$, $4f 5p^{6} 6p$,\\ $4f 5p^{6} 5d$\end{tabular} & 303 & 2811 & 2811\\ 
Nd VI & \begin{tabular}{l} ${\bf 4f 5p^{6}}$, $4f^{2} 5p^{5}$, $5p^{6} 7p$, $5p^{6} 6s$, $5p^{6} 6p$,\\ $5p^{6} 5d$\end{tabular} & 78 & 96 & 96\\ 
Nd VII & \begin{tabular}{l} ${\bf 5p^{6}}$, $5p^{4} 4f^{2}$, $5p^{5} 4f$, $5p^{5} 6s$, $5p^{5} 6p$,\\ $5p^{5} 5d$\end{tabular} & 210 & 1274 & 1274\\ 
Nd VIII & \begin{tabular}{l} ${\bf 5p^{5}}$, $4f 5p^{4}$, $4f^{2} 5p^{3}$, $4f 5p^{3} 6s$, $4f 5p^{3} 6p$,\\ $4f 5p^{3} 5d$\end{tabular} & 926 & 96706 & 96706\\ 
Nd IX & \begin{tabular}{l} ${\bf 4f 5p^{3}}$, $5p^{4}$, $5p^{2} 4f^{2}$, $4f 5p^{2} 6s$, $4f 5p^{2} 6p$,\\ $4f 5p^{2} 5d$\end{tabular} & 730 & 59206 & 59206\\ 
Nd X & \begin{tabular}{l} ${\bf 4f 5p^{2}}$, $5p^{3}$, $4f^{2} 5p$, $4f 5p 6s$, $4f 5p 6p$,\\ $4f 5p 5d$\end{tabular} & 312 & 11561 & 11561\\ 
Nd XI & \begin{tabular}{l} ${\bf 5p 4f}$, $5p^{2}$, $4f^{2}$, $5p 6s$, $5p 6p$,\\ $5p 5d$\end{tabular} & 56 & 337 & 337\\ 
\hline
\\
Sm I & \begin{tabular}{l} ${\bf 4f^{6} 6s^{2}}$, ${\bf 4f^{6} 6s 5d}$, ${\bf 4f^{6} 6s 6p}$, ${\bf 4f^{6} 6s 7s}$, $4f^{5} 5d 6s^{2}$,\\ $4f^{5} 5d^{2} 6s$\end{tabular} & 28221 & 43903717 & 54329\\ 
Sm II & \begin{tabular}{l} ${\bf 4f^{6} 6s}$, ${\bf 4f^{7}}$, $4f^{6} 5d$, $4f^{6} 6p$, $4f^{5} 5d 6s$\end{tabular} & 9030 & 5842197 & 1459980\\ 
Sm III & \begin{tabular}{l} ${\bf 4f^{6}}$, $4f^{5} 5d$, $4f^{5} 6s$, $4f^{5} 6p$\end{tabular} & 3737 & 1045697 & 985731\\ 
Sm IV & \begin{tabular}{l} ${\bf 4f^{5}}$, $4f^{4} 5d$, $4f^{4} 6s$, $4f^{4} 6p$\end{tabular} & 1994 & 320633 & 320091\\ 
Sm V & \begin{tabular}{l} ${\bf 4f^{4} 5p^{6}}$, $4f^{5} 5p^{5}$, $4f^{3} 5p^{6} 6s$, $4f^{3} 5p^{6} 6p$, $4f^{3} 5p^{6} 5d$,\\ $4f^{3} 5p^{6} 7s$\end{tabular} & 2067 & 283093 & 283093\\ 
Sm VI & \begin{tabular}{l} ${\bf 4f^{3} 5p^{6}}$, $4f^{4} 5p^{5}$, $4f^{3} 5p^{5} 6s$, $4f^{3} 5p^{5} 6p$, $4f^{3} 5p^{5} 5d$,\\ $4f^{3} 5p^{5} 7s$\end{tabular} & 5230 & 2288568 & 2288568\\ 
Sm VII & \begin{tabular}{l} ${\bf 4f^{3} 5p^{5}}$, $4f^{2} 5p^{6}$, $4f^{3} 5p^{4} 6s$, $4f^{3} 5p^{4} 6p$, $4f^{3} 5p^{4} 5d$,\\ $4f^{3} 5p^{4} 7s$\end{tabular} & 11589 & 9998002 & 9998002\\ 
Sm VIII & \begin{tabular}{l} ${\bf 4f^{3} 5p^{4}}$, $4f^{2} 5p^{5}$, $4f^{3} 5p^{3} 6s$, $4f^{3} 5p^{3} 6p$, $4f^{3} 5p^{3} 5d$,\\ $4f^{3} 5p^{3} 7s$\end{tabular} & 15567 & 18619221 & 18619221\\ 
Sm IX & \begin{tabular}{l} ${\bf 4f^{3} 5p^{3}}$, $4f^{2} 5p^{4}$, $4f^{3} 5p^{2} 6s$, $4f^{3} 5p^{2} 6p$, $4f^{3} 5p^{2} 5d$,\\ $4f^{3} 5p^{2} 7s$\end{tabular} & 12293 & 11835344 & 11835344\\ 
Sm X & \begin{tabular}{l} ${\bf 4f^{3} 5p^{2}}$, $4f^{2} 5p^{3}$, $4f^{3} 5p 6s$, $4f^{3} 5p 6p$, $4f^{3} 5p 5d$,\\ $4f^{3} 5p 7s$\end{tabular} & 5388 & 2497192 & 2497192\\ 
Sm XI & \begin{tabular}{l} ${\bf 4f^{3} 5p}$, $4f^{2} 5p^{2}$, $4f^{3} 6s$, $4f^{3} 6p$, $4f^{3} 5d$,\\ $4f^{3} 7s$\end{tabular} & 1205 & 130432 & 130432\\
\\
Eu I & \begin{tabular}{l} ${\bf 4f^{7} 6s^{2}}$, $4f^{7} 5d 6s$, $4f^{7} 6s 6p$, $4f^{6} 5d 6s^{2}$, $4f^{7} 5d 6p$,\\ $4f^{7} 6s 7s$, $4f^{6} 5d^{2} 6s$, $4f^{7} 5d^{2}$, $4f^{7} 6s 7p$, $4f^{7} 6s 6d$,\\ $4f^{7} 6s 8s$, $4f^{7} 6s 5f$, $4f^{7} 6s 8p$, $4f^{7} 6s 7d$, $4f^{7} 6p^{2}$\end{tabular} & 103229 & 741430825 & 4101\\ 
Eu II & \begin{tabular}{l} ${\bf 4f^{7} 6s}$, $4f^{7} 5d$, $4f^{7} 6p$, $4f^{6} 5d 6s$, $4f^{6} 5d^{2}$\end{tabular} & 22973 & 21396542 & 910949\\ 
Eu III & \begin{tabular}{l} ${\bf 4f^{7}}$, $4f^{6} 5d$, $4f^{6} 6s$, $4f^{6} 6p$\end{tabular} & 5323 & 2073702 & 1651778\\ 
Eu IV & \begin{tabular}{l} ${\bf 4f^{6}}$, $4f^{5} 5d$, $4f^{5} 6s$, $4f^{5} 6p$\end{tabular} & 3737 & 1045697 & 1045697\\ 
Eu V & \begin{tabular}{l} ${\bf 4f^{5} 5p^{6}}$, $4f^{6} 5p^{5}$, $4f^{4} 5p^{6} 6s$, $4f^{4} 5p^{6} 6p$, $4f^{4} 5p^{6} 5d$,\\ $4f^{4} 5p^{6} 7s$\end{tabular} & 3897 & 1140035 & 1137991\\ 
Eu VI & \begin{tabular}{l} ${\bf 4f^{4} 5p^{6}}$, $4f^{5} 5p^{5}$, $4f^{4} 5p^{5} 6s$, $4f^{4} 5p^{5} 6p$, $4f^{4} 5p^{5} 5d$,\\ $4f^{4} 5p^{5} 7s$\end{tabular} & 13065 & 12823350 & 12819025\\ 
Eu VII & \begin{tabular}{l} ${\bf 4f^{4} 5p^{5}}$, $4f^{3} 5p^{6}$, $4f^{4} 5p^{4} 6s$, $4f^{4} 5p^{4} 6p$, $4f^{4} 5p^{4} 5d$,\\ $4f^{4} 5p^{4} 7s$\end{tabular} & 29465 & 60643899 & 60636013\\ 
Eu VIII & \begin{tabular}{l} ${\bf 4f^{4} 5p^{4}}$, $4f^{3} 5p^{5}$, $4f^{4} 5p^{3} 6s$, $4f^{4} 5p^{3} 6p$, $4f^{4} 5p^{3} 5d$,\\ $4f^{4} 5p^{3} 7s$\end{tabular} & 40241 & 113753012 & 113745357\\ 
Eu IX & \begin{tabular}{l} ${\bf 4f^{4} 5p^{3}}$, $4f^{3} 5p^{4}$, $4f^{4} 5p^{2} 6s$, $4f^{4} 5p^{2} 6p$,\\ $4f^{4} 5p^{2} 5d$, $4f^{4} 5p^{2} 7s$\end{tabular} & 31393 & 73355941 & 73355941\\ 
Eu X & \begin{tabular}{l} ${\bf 4f^{4} 5p^{2}}$, $4f^{3} 5p^{3}$, $4f^{3} 5p^{2} 6s$, $4f^{3} 5p 6s^{2}$, $4f^{4} 5p 6s$,\\ $4f^{4} 5p 6p$, $4f^{4} 5p 5d$, $4f^{4} 5p 7s$\end{tabular} & 15515 & 18807502 & 18787178\\ 
Eu XI & \begin{tabular}{l} ${\bf 4f^{4} 5p}$, $4f^{3} 5p^{2}$, $4f^{4} 6s$, $4f^{4} 6p$, $4f^{4} 5d$,\\ $4f^{4} 7s$\end{tabular} & 3204 & 853861 & 853861\\
 \label{tab:confg}
\end{longtable}
\end{center}

\setcounter{figure}{0}
\section{Convergence test for opacity}\label{appendix:B}
In this section, we explore the completeness of our new atomic data for calculating the opacity.
For this purpose, we perform the convergence test on the opacity by using the atomic data corresponding to only a subset of configurations for IX ions of lanthanides (\ar{fig:conv_la}).
We choose IX ion because this is one of the major contributors to the opacity at high temperature ($\sim 70000$ K) 
due to the highly dense energy levels. \ar{fig:conv_la} shows the distribution of the energy levels for individual configurations (left)
and the Planck mean opacities calculated using a subset of atomic data (right).
The opacity using the default configuration set, i.e., all the configurations mentioned in \ar{tab:confg}, is represented by the thick black curve.

The opacities for Eu IX remain unaffected as long as the energy levels belonging to the configurations up to $4f^4\,5p^2\,6p$ are included. This implies that the transitions to or from the energy levels belonging to $4f^4 \, 5p^2\,7s$ do not have a significant impact on the Planck mean opacity, mostly due to the negligible population in these relatively high energy levels (e.g., the energy levels belonging to $4f^4 \, 5p^2\,7s$ are $> 80$ eV, left panel of \ar{fig:conv_la}). However, further removal of energy levels introduces about a factor of $\sim 2$ difference in the opacity for Eu IX.
Similar trends are found for Sm IX.
On the other hand, for the case of Nd IX, the opacities are affected up to a factor of $\sim 2$ if the energy levels belonging to the configurations up to $4f\,5p^2\,6p$ are not included. 
This is due to relatively low energy levels of the excited configurations in Nd. 
However, as the number density of the levels is not extremely high for Nd IX, the opacity itself is small, as compared with Eu IX and Sm IX.
Therefore, we conclude that our opacities are mostly converged, and further addition of excited configurations will have a negligible effect on opacity.

\begin{figure*}[t]
  \begin{tabular}{c}

   \begin{minipage}{0.5\hsize}
      \begin{center}
        \includegraphics[width=\linewidth]{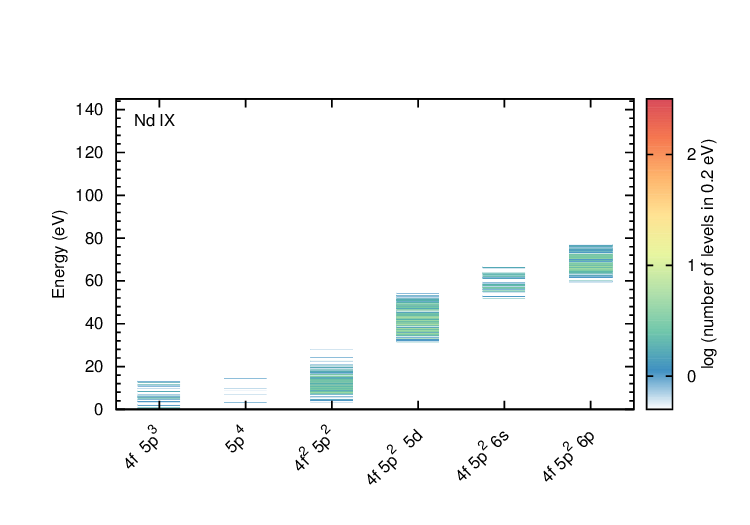}
      \end{center}
    \end{minipage}
    
     \begin{minipage}{0.5\hsize}
      \begin{center}

        \includegraphics[width=\linewidth]{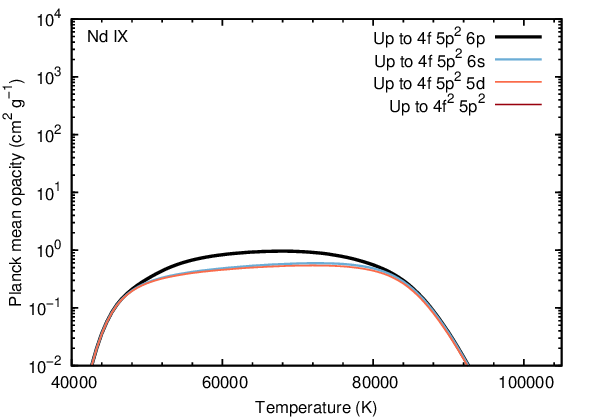}
      \end{center}
      \end{minipage}

  \\
   \begin{minipage}{0.5\hsize}
      \begin{center}
        \includegraphics[width=\linewidth]{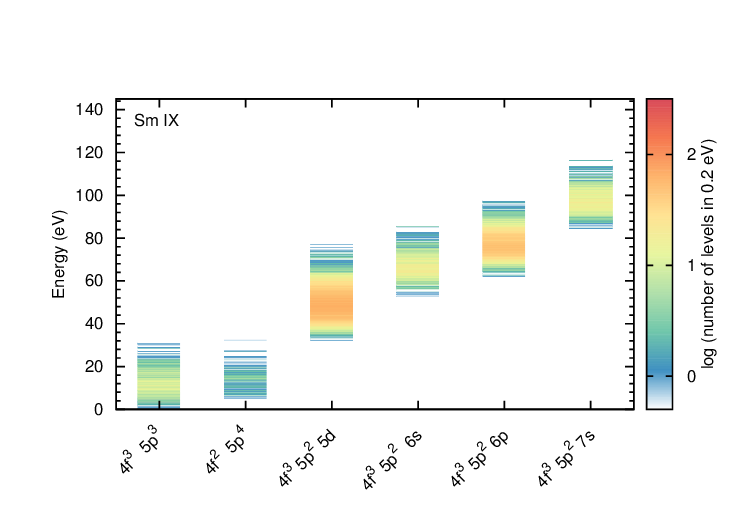}
      \end{center}
    \end{minipage}
    
     \begin{minipage}{0.5\hsize}
      \begin{center}

        \includegraphics[width=\linewidth]{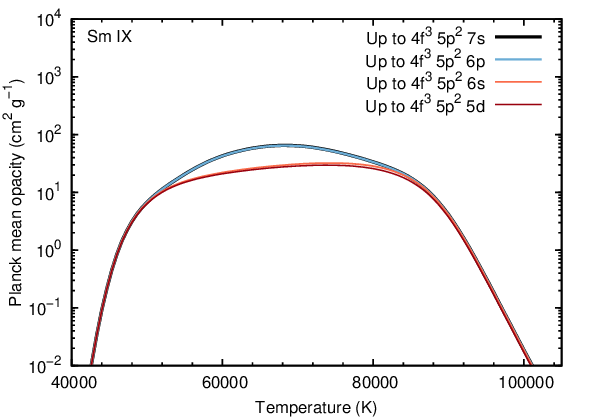}
      \end{center}
      \end{minipage}

  \\

   \begin{minipage}{0.5\hsize}
      \begin{center}
        \includegraphics[width=\linewidth]{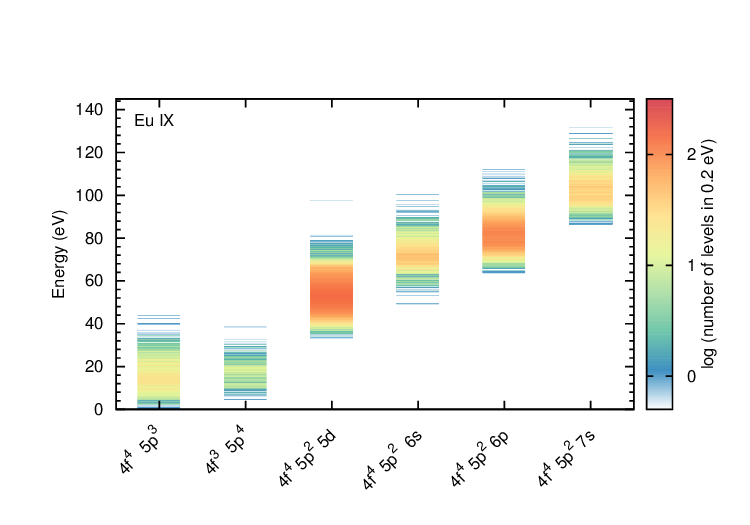}
      \end{center}
    \end{minipage}
    
     \begin{minipage}{0.5\hsize}
      \begin{center}

        \includegraphics[width=\linewidth]{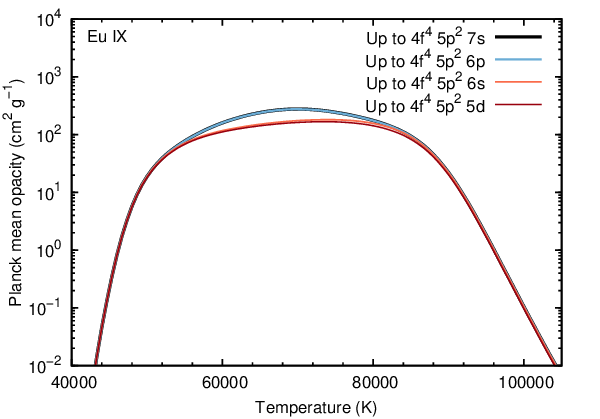}
      \end{center}
      \end{minipage}

\end{tabular}
  \caption{The distribution of the energy levels within the energy bin of 0.2 eV for the individual configurations
  as obtained from HULLAC (left panel) and the corresponding Planck mean opacities (right panel) for Nd IX, Sm IX, and Eu IX ions
  (top to the bottom panels).
  Different colors of the curves in right panel correspond to the opacities calculated by including different subsets of the configurations as shown in the legend.}
  \label{fig:conv_la}
\end{figure*}
\setcounter{figure}{0}
\section{Scheme for reduced linelist}\label{appendix:C}
For the highly ionized lanthanides, the number of lines is exceptionally high as described in \ar{sec:atomic}.
The numbers of transition for one ion reaches $\sim 0.1$ billion in some cases (\ar{tab:confg}).
With such a large linelist, performing radiative transfer simulation becomes infeasible.
For the purpose of making the radiative transfer simulation possible, we create a reduced linelist by randomly choosing
a single line out of $n_{\rm sample}$ number of lines.
If the reduced linelist preserves the statistical properties of the original linelist,
i.e., if the statistical distributions of the transition wavelengths, radiative transition probabilities,
and the statistical weights of the energy levels are preserved, the opacity with the reduced linelist
can approximately reproduce the original result, provided the contribution from the selected lines is enhanced by a factor of
$n_{\rm sample}$ (see \ar{eqn:kexp_app}).

We test our scheme by calculating the expansion opacity with the reduced linelist as:
\begin{equation}\label{eqn:kexp_app}
    \kappa_{\rm{exp}}(\lambda) = \bigg  [\frac{1}{\rho ct}\sum_{l}\frac{\lambda_{l}}{\Delta \lambda}(1 -e^{-\tau_{l}}) \bigg ]  \times n_{\rm sample}
\end{equation}
Now the summation is taken over the reduced linelist.
We adopt $n_{\rm sample} = 1000$ for the linelist for lanthanides, i.e., we choose 1 out of 1000 lines to reduce the linelist to $0.1\%$ of the full linelist.
The expansion opacity spectra show a sound agreement with the original result,
although the opacity becomes noisier at longer wavelengths because of random sampling
(left panel of \ar{fig:rand1}).
The Planck mean opacity shows a perfect match (right panel of \ar{fig:rand1}).
Therefore, we conclude that our scheme of the reduced linelist preserves the statistical
properties of the original linelist, and thus, can be used for the opacity calculation in
radiative transfer simulations.
Since the opacity spectra becomes noisier, the detailed features of the energy spectra calculated using the reduced linelist are affected.
However, the effect on the overall bolometric lightcurve and the broad-band magnitudes are not significant because they are derived
by integrating the energy spectra over a wide wavelength range.


\begin{figure*}[t]
  \begin{tabular}{c}
 
   \begin{minipage}{0.5\hsize}
      \begin{center}
        \includegraphics[width=\linewidth]{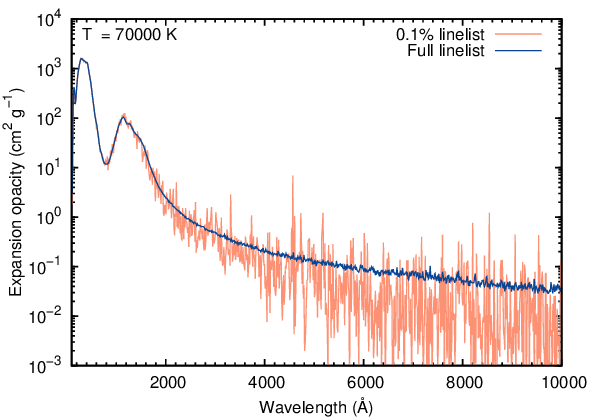}
      \end{center}
    \end{minipage}
    
     \begin{minipage}{0.5\hsize}
      \begin{center}

        \includegraphics[width=\linewidth]{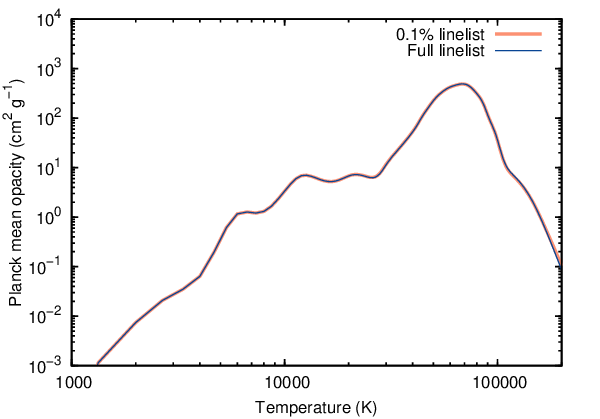}
      \end{center}
      \end{minipage}
      
\end{tabular}
  \caption{Comparison between the expansion (left panel) and Planck mean opacity (right panel) calculated
    by using the reduced and the original linelist for Eu.}
  \label{fig:rand1}
\end{figure*}


\end{document}